\documentclass{epl}

\title{Molecular Motor with a Build-In Escapement Device}
\shorttitle{Molecular Motor}
\author{G.Oshanin\inst{1,2} \and J.Klafter\inst{3} \and
M.Urbakh\inst{3}}
\institute{
  \inst{1} Laboratoire de Physique Th\'eorique des Liquides,
Universit\'e Paris 6, Tour 16, 4 place Jussieu, 75252 Paris
Cedex 05, France\\
  \inst{2} Max-Planck-Institut f\"ur Metallforschung, Heisenbergstr. 3,
D-70569 Stuttgart, Germany, \and
Institut f\"ur Theoretische und Angewandte Physik,
Universit\"at Stuttgart, Pfaffenwaldring 57, D-70569 Stuttgart,
Germany\\
  \inst{3} School of Chemistry, Tel Aviv University, 
69978 Tel Aviv, Israel
}

\begin{document}

\maketitle

\begin{abstract}
We study dynamics of a classical particle in a
one-dimensional  potential, which is composed of two
periodic components, 
that are time-independent, 
have equal amplitudes and
periodicities. One of them is externally driven 
by a random force  and thus
performs a
diffusive-type motion  with 
respect to the other. 
We demonstrate that here, under certain conditions, 
the particle may move
unidirectionally with a $constant$ velocity, despite the fact that
the random force averages out to zero.
We show that
the physical 
mechanism underlying such a phenomenon resembles the work 
of an escapement-type device in watches;
upon reaching certain level,  random 
fluctuations 
exercise a locking function
creating the points
of irreversibility in particle's trajectories 
such that 
the particle gets uncompensated
displacements. Repeated (randomly) in each cycle,
this process ultimately
results
in a random ballistic-type motion. 
In the overdamped limit, 
we  work out simple analytical 
estimates for the particle's terminal 
velocity. Our analytical results are in a very good agreement
with the Monte Carlo data.
\end{abstract}

\section{Introduction}

A long standing challenge
has been a problem of
how to 
rectify unbiased random or 
periodic 
fluctuations into 
directional motion.  
For microscopic fluctuations,
induced, e.g., 
by Brownian noise, this question 
has been 
debated 
since the inception
of the random walk theory, and 
several important concepts have been worked
out 
\cite{maxwell,smoluchowski} (see also Ref.\cite{Feynman/Leighton/Sands:1963}).
Indeed, it has been established
already by Curie \cite{Curie:1894} more than a hundred years ago
that although 
a violation of
the $x \rightarrow -x$ symmetry is 
not sufficient to cause a net directional
transport of
a particle subject to a spatially 
asymmetric but on large scale 
homogeneous
potential, 
the additional breaking of 
time reversal $t \rightarrow -t$ symmetry, (e.g., due to dissipative
processes), may lead to a macroscopic net velocity, such that here
directed motion can result 
in the absence of any external net force. 
These systems,
known currently as thermal ratchets 
\cite{Feynman/Leighton/Sands:1963}, 
have been subject of intensive research, both theoretical
(see, e.g. Refs.\cite{Ajdari/Prost:1993,Magnasco:1993,Astumian/Bier:1994,%
Duke:1995,Astumian:1997,Dialynas/Lindenberg/Tsironis:1997,%
Sokolov:1998+99,%
Fisher/Kolomeisky:1999,
Lipowski:2000,Porto/Urbakh/Klafter:2000,Popescu,Lipowski/Klumpp/Nieuwenhuizen:2001}) 
and experimental (see, e.g. 
Refs.\cite{Rousselet/Salome/Ajdari/Prost:1994,%
Faucheux/Bourdieu/Kaplan/Libchaber:1995,Gorre/Ioannidis/Silberzan:1996,
Kettner/Reimann/Haenggi/Mueller:2000}). 
Advancements in this field have been summarized in
Refs.\cite{Juelicher/Ajdari/Prost:1997,Reimann/Hanggi:2002,Reimann:2002}.

On the macroscopic scale, the same
problem 
of obtaining a useful work from
random or periodic 
perturbations has been 
known for many centuries and this task
has been indeed accomplished in many instances; 
to name but a few, we mention
water- or windmills, or watches. 
In the latter case, 
the watch makers had to create a mechanism capable
to convert
the raw power of the driving 
force into regular and uniform impulses, which
was realized 
by inventing various 
so-called $escapement$ devices 
(see, e.g., Ref.\cite{watch}). 
Such a devise
is most often a sort of a shaft 
or an arm carrying two tongues - pallets, which
alternately engage with the teeth of a crown-wheel. The pallets follow
the oscillating
motion of the controller - the balance-wheel or a pendulum,
and in each cycle of the controller the crown-wheel turns freely
only when both pallets are out of contact with it.
Upon contacts, the pallets
provide impulses 
to the crown-wheel,
(which is necessary 
to keep the controller from drifting to a halt), and moreover,
perform a locking function stopping the train of wheels 
until the swing of the controller
brings round the next period of release.

In this Letter we 
exploite the concept of such 
an escapement device to build a molecular motor
capable of doing a "useful work"  under the action of
random forces which average out to zero. The system we consider consists
of a classical particle in a
one-dimensional (1D) two-wave potential, which is is composed of two
periodic in space, identic time-independent components. 
One of them is externally driven 
by a random force  and thus
performs a random
translational motion  with 
respect to the other. 
This model has been previously proposed  in
Ref.\cite{Porto/Urbakh/Klafter:2000} and it has been discovered
that under certain conditions and for constant or periodic in time
external
driving force,
the particle may perform
totally directed motion with a constant velocity.
Here we demonstrate that in such a system the particle may move
$unidirectionally$ with a constant velocity even in situations
when the $direction$ of the driving 
force (the controller) 
fluctuates randomly in time.
We show that
the physical 
mechanism underlying such a phenomenon is indeed 
that of an escapement-type device -
upon reaching certain levels,  random 
fluctuations 
exercise a locking function
creating the points
of irreversibility in particle's trajectories 
such that, (despite the fact that
the fluctuations average to zero),
the particle gets uncompensated
displacements. Repeated (randomly) in each cycle,
this process ultimately
results
in a ballistic-type motion. 
Focussing on the overdamped limit, 
we use this physical picture to map the original
system onto  a Brownian motion (BM) process
on a hierarchy of disconnected intervals. This allows us
to work out simple analytical 
estimates for the particle's terminal 
velocity, which are in a very good agreement
with Monte Carlo data.

\section{The Model}

Following Ref.\cite{Porto/Urbakh/Klafter:2000}, we 
consider a simple piece-wise continuous potential 
\begin{equation}\label{eq:ratchet}
\Pi(x) \equiv \Pi_0 \left\{
\begin{array}{ll}
\displaystyle -1 + 2 \frac{x}{\xi} &
\mbox{if $\displaystyle x \le \xi$}\\[5mm]
\displaystyle 1 - 2 \frac{x- \xi}{1 - \xi} &
\mbox{if $\displaystyle x > \xi$}
\end{array}\right. \quad,
\end{equation}
where 
the parameter $\xi \in (0, 1 )$
determines the asymmetry of the potential, 
with $\xi = 1/2$ corresponding to the symmetric
case. We will constrain our analysis here to the case $\xi < 1/2$.
Note that
it has a periodicity $b = 1$, so
that $\Pi(x+1) = \Pi(x)$ $\forall x$, 
an amplitude $\Pi_0 = \max \Pi(x) = -\min
\Pi(x)$, and one minimum is located at $x = 0$, 
i.e.\ $\Pi(0) = -\Pi_0$. 
%Note also that arbitrary periodicity $b$ can be introduced
%and can be
%readily restored in our final results. 

Now, the total potential $V(x,\gamma)$ is the sum
 $V(x,\gamma) \equiv \Pi(x) + \Pi(x-\gamma)$, 
where $\gamma$ defines the
external translation. Because of the periodicity of the potentials $\Pi(x)$, the
potential $V(x,\gamma)$ is periodic 
in both arguments, so that $V(x,\gamma+1) =
V(x,\gamma)$ $\forall \gamma$.
In this potential landscape, the
equation of motion of a particle of mass $m$ in a 1D medium with damping $\eta$
reads:
\begin{equation}\label{eq:motion}
m \ddot{x}_t + \eta \dot{x}_t + \frac{\partial V(x_t,\gamma)}{\partial x_t} = 0,
\end{equation}
$x_t$ being the particle's trajectory. 
We note that
energy is steadily pumped 
into the system through
$\gamma$, and is
dissipated when $\eta > 0$, which prevents
particle's  detachement.

Finally, we specify the properties 
of $\gamma$. 
We suppose
that $\gamma$ obeys the Langevin equation, 
$\dot{\gamma} = f_t$,
where 
$f_t$ is a random Gaussian, delta-correlated force with moments
\begin{equation}
\label{force}
\overline{f_t} = f_0, \;\;\; \overline{f_t f_{t'}} = f_0^2 + 2 D \delta(t - t'),
\end{equation}
while
the overbar denotes
averaging 
over thermal histories. 
We will focus here mostly 
on the case 
$f_0 = 0$, such that 
$\gamma$ is just a trajectory
of a symmetric 1D BM. The case 
when $f_0 \neq
0$ or 
when 
$\gamma$ represents $anomalous$ diffusion
will be briefly discussed at the end of the paper.

\section{Particles Dynamics in the Overdamped Limit}

We will restrict ourselves
to the limit of  an overdamped motion,
$\eta/[(2 \pi) \; \sqrt{m \Pi_0}] \gg 1$.
As shown in Ref.\cite{Porto/Urbakh/Klafter:2000}, in this limit
dynamics of the particle is  governed entirely 
by the evolution of
different minima of the total potential $V(x,\gamma)$. 
Consequently, in order to obtain  the 
trajectory $x_t$ and the
average velocity $V = \overline{\dot{x}_t}$,
 it suffices to study the time evolution of 
positions of these minima. 
This process has been amply discussed in 
Ref.\cite{Porto/Urbakh/Klafter:2000}
and here we will merely outline the main conclusions.

According to  Ref.\cite{Porto/Urbakh/Klafter:2000},
the particle's dynamics proceeds as follows: 
the total potential $V(x,\gamma)$ 
possesses a set of minima
and position of each minima changes in time as the translation $\gamma$ evolves.
The particle, located at $t = 0$ at the first minimum 
simply follows the motion of this minimum up 
to a certain moment 
of time, or 
more precisely, a 
time moment when this minimum reaches a certain point
$x = \tilde{x}$. 
These $\tilde{x}$-points
 are the points of instability ${\cal I}$ in the $(x,\gamma)$ plane, where 
the corresponding local minimum of the potential $V(x,\gamma)$
ceases to exist, and which emerge 
due to the asymmetry of $\Pi(x)$. 
Further on, when the minimum disappears, a particle located at such
a point performs an irreversible motion jumping to one of two 
neighboring minima, which still exist. 
Since 
the potential in the vicinity 
of each $\tilde{x}$ may have both a downward and an upward 
bending, the jumps may be performed  in both 
the leftward and the rightward directions. Depending whether
$\xi < 1/2$ or $\xi > 1/2$, the leftward or rightward jumps 
may result in uncompensated or compensated displacements.

\section{Evolution of the effective translation}

We now put the particle's dynamic rules deduced in 
Ref.\cite{Porto/Urbakh/Klafter:2000}
into a mathematical framework, amenable 
to analytical analysis, 
and introduce an auxiliary stochastic process $\gamma_t$. 
We will call $\gamma_t$ as an "effective translation",
since it follows, apart of some special points (to be specified below),
the evolution of the potential minima visited  
by the particle
and keeps track of both the translation $\gamma$ and the effect
of the irreversibility points
 which induce particle's leftward and rightward jumps to neighboring
minima when the local minimum seizes to exist. We note also
that the particle's actual trajectory $x_t = \gamma_t$ almost everywhere, except for
some special points (to be specified below).

The effective translation $\gamma_t$ obeys
the following Smoluchowski-Feynman ratchet form:
\begin{equation}
\label{L}
\dot{\gamma}_t = \hat{\cal L}\left(\gamma_t\right) + f_t,
\end{equation}
where $f_t$ are random forces defined in eq.(\ref{force}),
while $\hat{\cal L}\left(\gamma_t\right)$ is a 
local operator, which may be thought of as a space-derivative
of some "effective" potential.

This operator equals 
zero everywhere except for the set of 
special points $\gamma_t = - K + \xi$ 
and $\gamma_t = - K + 3/2$, $K = 1, 2, \ldots$. 
At these special points, the action of the operator
$\hat{\cal L}\left(\gamma_t\right)$ is as follows:
When  $\gamma_t$ reaches 
$\gamma_t = - K + 3/2$, (which we call the "reflection point"),  
the operator $\hat{\cal L}\left(\gamma_t\right)$ changes 
$\gamma_t \to \gamma_t - 1$, (i.e.  
shifts its value from  
$\gamma_t = - K + 3/2$ to  $\gamma_t = - K + 1/2$).  
On the other hand, when  $\gamma_t$ hits
$\gamma_t = - K + \xi$, the operator $\hat{\cal L}\left(\gamma_t\right)$ 
shifts the 
position of the reflection point 
from  $- K + 3/2$ to $- K + 1/2$, thus performing a "locking" function - $\gamma_t$
can not now return
below the point $-K + 3/2$. 

\begin{figure}[ht]
\twofigures[width=7cm, angle=0]{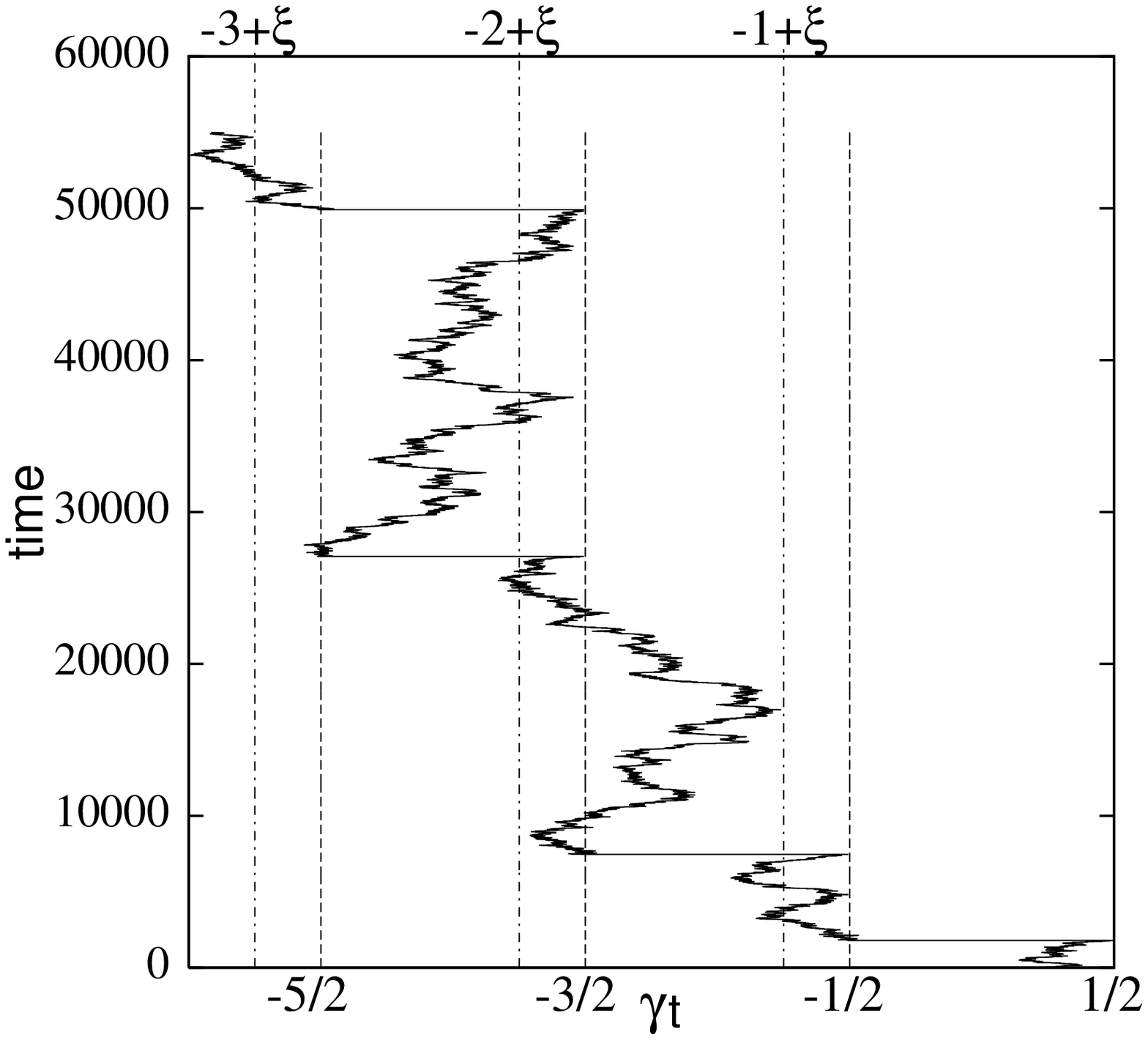}{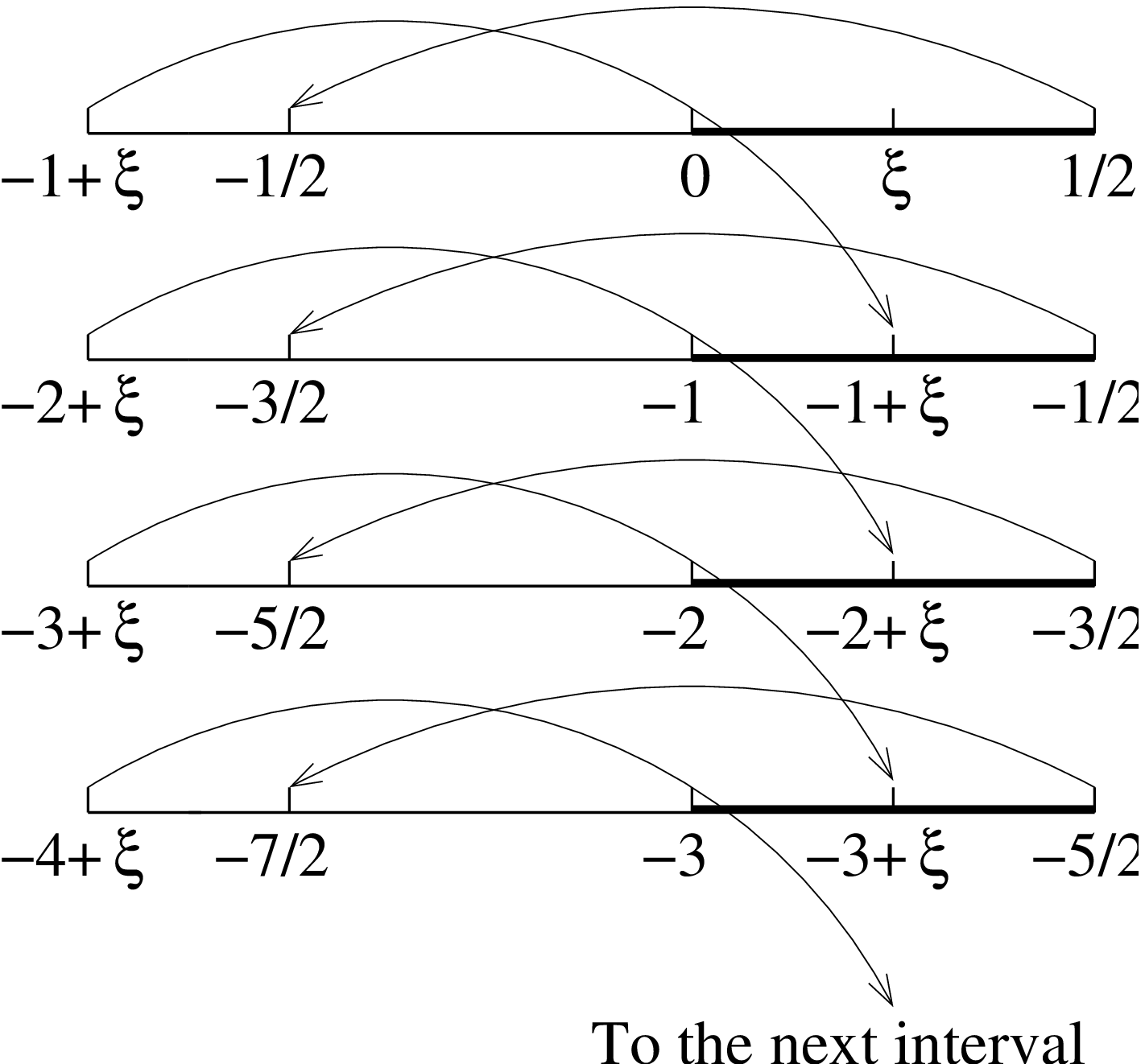}
\caption{{\small  
Typical realization of the effective translation $\gamma_t$. The dashed lines 
denote the "reflection points" $x=-K + 3/2$ of the trajectory $\gamma_t$, while 
the dash-dotted lines - the points $-K + \xi$. In simulations, we took $\xi = 1/4$,
the lattice 
spacing $\delta \gamma = 2 \; \times \; 10^{-2}$ and 
the characteristic jump time $\tau = 1$; hence, 
$D = (\delta \gamma)^2/2 \tau =  2 \; \times \; 10^{-4}$. }}
\label{Fig1}
\caption{{\small  
Time evolution of $\gamma_t$ viewed as a BM on a
hierarchy of disconnected intervals.}}
\label{Fig2}
\end{figure}
In Fig.1 we depict a typical 
realization of the process $\gamma_t$ for $f_0 = 0$.
 Note that because of $\hat{\cal L}\left(\gamma_t\right)$,  $\gamma_t$
experiences an effective drift in the negative direction 
(for $\xi < 1/2$); for
$\hat{\cal L}\left(\gamma_t\right) \equiv 0$ 
one should, of course, regain a standard result
$\overline{\gamma_t} = \overline{\gamma} \equiv 0$ and $\overline{\gamma^2_t} =
\overline{\gamma^2} \equiv 2 D t$.

Note now that   
$\gamma_t$ can be viewed from a different perspective;
 namely, $\gamma_t$ can be regarded 
as a BM 
on a 
hierarchical lattice composed of a semi-infinite set of 
disconnected intervals of length $3/2 - \xi$ (see Fig.2). 
The process starts at $t = 0$ 
at the origin of the first interval ($K = 1$) and evolves freely until it either 
hits the right-hand-side (RHS) boundary $\gamma_t = 1/2$ (the reflection point), 
in which case 
it gets transferred instantaneously at position $\gamma_t = - 1/2$ and 
continues its motion 
on the interval $K = 1$, or reaches the left-hand-side (LHS) 
boundary $\gamma_t = - 1 + \xi$ - the "exit point" 
and gets irreversibly
transferred to the second ($K = 2$) interval. In the second, and etc, interval
$\gamma_t$  evolves according to the same rules. 

The particle's trajectory  $x_t$ 
follows the process $\gamma_t$, 
i.e. $x_t = \gamma_t$,  
except for two 
details: a) when $\gamma_t \in [- K + 1, - K + 3/2]$, $K = 1, 2, \ldots$, 
(thick lines in Fig.2), 
$x_t$ sticks to the left boundary of these intervals,  such that the trajectories $x_t$
are saltatory with the random pausing times
corresponding to the times
spent 
by $\gamma_t$ 
in the intervals $[- K + 1, - K + 3/2]$, $K = 1, 2, \ldots$, (see Fig.1).
b) the event when $\gamma_t$ hits the LHS
boundary of the $K$-th interval and gets 
instantaneously transferred to the next interval 
(i.e. $\gamma_t$ does not have a discontinuity) 
corresponds to the event when the particle 
makes a jump from $x_t = - K +\xi$ to $x_t = - K$, i.e. 
at this point $x_t$ changes 
discontinuously. On the other hand, it is clear 
that such jumps 
do not contribute to the average particle velocity $ \overline{\dot{x}_t}$ 
and hence, $\overline{\dot{x}_t} = V = 
\overline{\dot{\gamma}_t}$.

\section{Particle's terminal velocity}

We turn next to the computation $V$ for  $f_0 = 0$. 
Evidently, even in 
this simplest case the map depicted in Fig.2 is too complex 
to be solved exactly. However, some simple analytical arguments can be 
proposed to obtain   very accurate estimates of  
 $V$. 
To do this, we first note that it is only when passing 
from the $K$-th interval 
to the $(K+1)$-th one, (which is displaced
on a unit distance on the $\gamma_t$-axis 
with respect to the previous one), 
the process $\gamma_t$ gains an uncompensated 
(negative) displacement
equal to the overlap distance 
of two consequitive intervals, i.e.,  $1/2 - \xi$. 
Consequently,  we may estimate $V$ as
$V = - (1/2 - \xi)/T$, where $T$ is the mean time 
which $\gamma_t$ "spends" within a given interval. 

\begin{figure}[ht]
\begin{center}
\twofigures[width=7cm, angle=0]{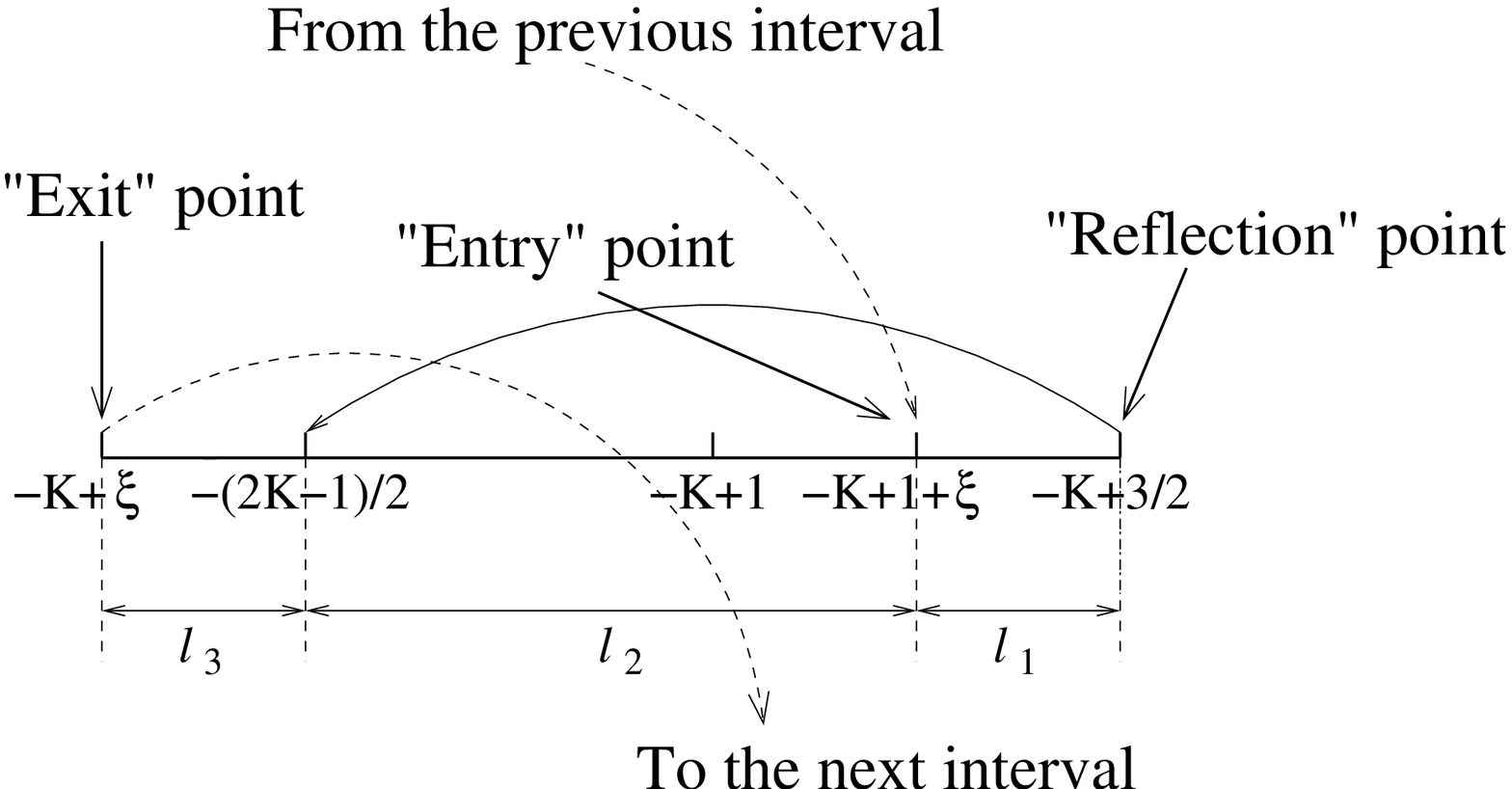}{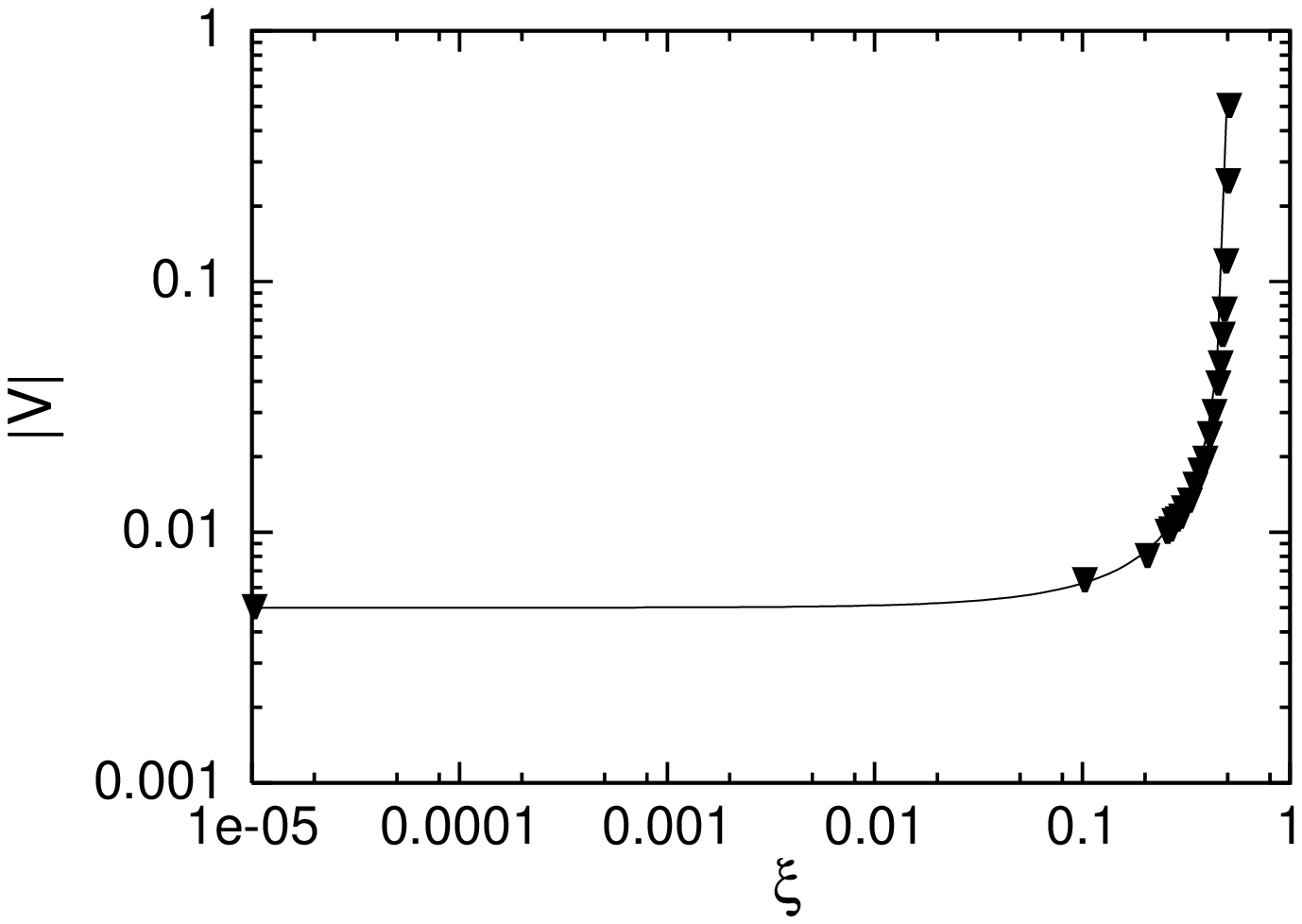}
\caption{{\small  
Time evolution of the effective external translation $\gamma_t$ on the $K$-th interval.
 }}
\label{Fig3}
\caption{{\small  The absolute value of $V$ versus $\xi$.
Solid line gives the analytical prediction in eq.(\ref{velocity}), while
the triangles depict Monte Carlo data. Note that we use, for notational
convenience, the log-log scale.
}}
\label{Fig4}
\end{center}
\end{figure}
To estimate 
$T$, 
let us consider how $\gamma_t$  evolves on a given interval;  
$T$ is, as a matter of fact, the mean time needed 
for  $\gamma_t$ to reach diffusively for the first time 
the LHS boundary of the interval - the "exit" point $- K + \xi$, 
starting from the "entry" point $-K + 1 + \xi$ (see Fig.3). 
We divide
next the $K$-th interval into three sub-intervals: 
$l_1 = l_3 = 1/2 - \xi$ and $l_2 = 1/2 + \xi$, Fig.3. 
Since $l_1 = l_3 \leq l_2$, ($\xi > 0$), 
for typical realizations 
of $f_t$, 
the following 
scenario should hold: $\gamma_t$ evolves randomly around the 
"entry" point and first hits the 
RHS boundary of the $K$-th interval passing 
thus through the sub-interval $l_1$; 
the mean time 
required for the first passage of $l_1$ is denoted as $T_1$.    
For BM with diffusion coefficient $D$, 
$T_1 = (1/2 - \xi)^2/2 D$ \cite{1}. 
Further on, after 
hitting the RHS boundary,  $\gamma_t$ gets reflected 
to the point $\gamma_t = - (2 K - 1)/2$ and 
evolves randomly around this point until it hits
the LHS boundary of the 
interval - the "exit" point. For symmetric BM, 
the time $T_3$ 
needed for the first passage through $l_3$ is equal to $T_1$.
Consequently, $T = T_1 + T_3 =   (1/2 - \xi)^2/D$ and 
the velocity reads
\begin{equation}
\label{velocity}
V = - 2 D \left(1 - 2 \xi\right)^{-1}
\end{equation}
In Fig.4 we compare our prediction in eq.(\ref{velocity}) and 
the results of Monte Carlo simulations, which 
shows that theoretical arguments presented
here capture the essential physics underlying the dynamics of the  
random map depicted in Fig.2 and, hence, of the process in 
eqs.(\ref{eq:motion}) and (\ref{force}).

Now, several comments on the result in eq.(\ref{velocity}) are in order.

(i) First, we note that  $V$, eq.(\ref{velocity}), 
diverges when $\xi \to 1/2$ 
(note, however, that  $ V \equiv 0$ for $\xi \equiv 1/2$), 
which is a seemingly counter-intuitive behavior. 
Such a behavior stems, as a matter of fact, 
from our definition of 
$\gamma_t$; namely, by writing our
 eq.(\ref{L}), we tacitly assume that both the 
effective step $\delta \gamma$ of the process $\gamma_t$ and the characteristic
jump time 
$\tau$ are infinitesimal variables, (while the ratio 
$D = (\delta \gamma)^2/2 \tau$ is supposed to be fixed and finite). 
On the other hand, for 
any real system $\delta \gamma$ and $\tau$ might be very small 
but nonetheless are both finite. 
For finite $\delta \gamma$ and $\tau$, 
$\gamma_t$ is a
symmetric hopping process
 on a lattice with spacing $\delta \gamma$ and with the characteristic
time  $\tau$.  Then, we 
have $T_1 = T_3 = \tau \Big(1 + L\Big) L$,
  where $L = (1/2 - \xi)/\delta \gamma$ is the number of
 elementary steps $\delta \gamma$ 
in the interval $(1/2 - \xi)$. 
This 
yields, in place of eq.(\ref{velocity}),
\begin{equation}
\label{velocityfin}
 V = - 2 D \left(2 \delta \gamma + 1 - 2 \xi\right)^{-1},
\end{equation}
where now $V$ 
tends to a finite value 
when $\xi \to 1/2$ (but still $V \equiv 0$ when $\xi \equiv 1/2$). 

(ii) Second, notice 
that $V$ in eq.(\ref{velocity})  is a monotoneously increasing function 
of $\xi$; that is, the
absolute value of $V$
is maximal when $\xi \to 1/2$ and minimal, $|V| = 
2 D$, for the strongest asymmetry, 
$\xi = 0$. 
We note that
such a behavior 
again stems from
the 
definition of $f_t$
as a Gaussian, \textit{delta-correlated} noise. 
On the other hand, since $f_t$ is influenced by some external
processes, it might be 
characterized, e.g., 
by correlations or be a non-local in space 
or in time (discontinuous) 
stochastic process. 
In both cases, the behavior of $V$ in
the presence of such an external force would be different of that predicted by
eqs.(\ref{velocity}) and
(\ref{velocityfin}).  

Consider now what will happen if $f_{t}$ represents the
so-called delta-correlated 
L\'evy  process (LP) with parameter $\mu$ 
(see an  exposition in excellent Ref.\cite{1} 
for more
details). 
We note here parenthetically 
that the case $\mu = 2$ corresponds to the usual 
Gaussian case, which yields conventional BM, while 
 the case $\mu = 1$ describes the so-called Cauchy process.

Now, what basically changes when we assume that $f_t$ is the LP,
is the form of the first passage times $T_1$ and $T_2$. Here, the
time required for the first passage of an interval of length $1/2 - \xi$ 
reads $T_1 = T_3 \sim (1/2 - \xi)^\mu$ \cite{1} and hence, 
$V \sim - (1 - 2 \xi)^{1 - \mu}$.
  
Consequently, we infer that $V$ will be
a monotoneously increasing function of the parameter $\xi$ 
only for the LP with $\mu > 1$, i.e. for the persistent processes (which have a finite
probability for return to the origin). In the bordeline case of $\mu = 1$ 
 velocity will be independent of the asymmetry parameter.
On contrary, for transient
(having zero probability of return to the origin) 
processes with $\mu < 1$, 
which are not
space-filling (fractal) 
and occupy the space in clustered or localized patches,
one would find that $V$ is a $decreasing$
function of $\xi$.

(iii) Finally, we consider the case when the bias is oriented in 
the positive direction, i.e. $f_0
\geq 0$. In this case, 
$\gamma_t$ is an asymmetric hopping process
on a 1D discrete lattice of spacing
$\delta \gamma$ and with transition
probabilities  $p$ ( $q$ ) of jumps 
in the positive  (negative) directions which 
obey
$p/q = \exp(\beta f_0 \delta \gamma)$, with $ p + q = 1$,
where $\beta$ denotes the reciprocal temperature.

Here,
 the symmetry $T_1 = T_3$, which exists for $f_0 = 0$,  is broken: when
passing through $l_1$ the process $\gamma_t$ follows the field,
while the passage through $l_3$ takes place  against it. 
Supposing
that $\delta \gamma$ and $\tau$ are both finite, we have
that here:
\begin{equation}
\label{T1}
T_{1,3} = \frac{\tau}{p - q} \left[\pm \frac{1 - 2 \xi}{2 \delta \gamma} \mp
\frac{\phi_i}{p-q}
\left(\left(\frac{p}{q}\right)^{\mp (L-1)} - 1\right)    \right], 
\end{equation}  
where the upper (lower) sign corresponds to index $"1"$ ($"3"$), while 
$\phi_1 = q$ and $\phi_3 = p$. 
Note that $T_1$ grows linearly with the sub-interval length $1/2 - \xi$,
while $T_3$ shows much stronger, $exponential$ interval-length dependence, and hence,
controls the overall time spent within a given interval.
Consequently, the particle's velocity in this case attains the form
\begin{equation}
\label{velocityf}
V  = - \frac{(1/2 - \xi) \sinh^2\Big(\beta f_0 \delta
\gamma/2\Big)}{\tau \cosh\Big(\beta f_0 \delta
\gamma/2\Big) \Big[ \cosh\Big(\beta f_0 (1 - 2 \xi + \delta \gamma)/2\Big) 
-  \cosh\Big(\beta f_0 \delta
\gamma/2\Big)\Big]},
\end{equation} 
which reduces in the diffusion limit to
the following result
\begin{equation}
\label{velocityfdif}
V = - \frac{(1 - 2 \xi) D \beta^2 f_0^2 }{8
\sinh^2\Big(\beta f_0 (1 - 2 \xi)/4\Big)}.
\end{equation} 
The salient feature of these results is that 
here the particle's drift
proceeds $against$ the applied field,  due to fluctuations 
in the process $\gamma_t$ - some
(exponentially small) number of trajectories which 
travel against the field.   

In conclusion,  we have studied dynamics of a classical
particle  in a
1D potential, composed of two
periodic  components, one of which 
is driven 
by an external random force. 
We have shown 
that in such a system the particle may move
$unidirectionally$ with a constant velocity 
even 
when the random driving 
force averages out to zero.
We have demonstrated that
the physical 
mechanism underlying such a behavior resembles the work
of the so-called escapement device, used by watch makers
to convert
the raw power of the driving 
force into uniform impulses; here, indeed,
upon reaching certain levels,  random forces 
lock the particle's motion
creating the points
of irreversibility,
such that
the particle gets uncompensated
displacements. Repeated (randomly) in each cycle,
this process ultimately
results
in a ballistic-type motion. 
Concentrating  on the overdamped limit, 
we have used this picture to map the original
system onto a BM  process
on a hierarchy of disconnected intervals, which allowed us
to present analytical estimates for the particle's  velocity.
Our analytical 
results are in a very good agreement
with Monte Carlo data. Extensions for systems with other than Gaussian fluctuations
and systems with global bias have also been presented.

\section{Acknowledgments}
The authors gratefully acknowledge helpful discussions with M.Porto. 
GO thanks the AvH Foundation for the financial 
support via the Bessel Research Award.

\end{document}